\documentclass[3p,twocolumn]{elsarticle}

\usepackage{graphicx}
\usepackage{subfig}
\usepackage{amssymb}
\usepackage{gensymb}
\usepackage{lineno}
\usepackage{multirow}
\usepackage{color}

\biboptions{comma,square,sort&compress}

\journal{Nuclear Instruments and Methods A}

\usepackage{cleveref}
\crefname{figure}{Fig.\@}{Figs.\@}
\crefname{equation}{Eq.\@}{Eqs.\@}
\crefname{section}{Sec.\@}{Secs.\@}
\crefname{table}{Table}{Tables}

\begin{document}


\begin{frontmatter}

\title{Design and Performance of a Lead Fluoride Detector as a Luminosity Monitor}

\author[inst1]{R.~P\'{e}rez Benito}
\author[inst1]{D.~Khaneft}
\author[inst2]{C.~O'Connor\corref{cor1}}
\cortext[cor1]{Corresponding Author}
\ead{colton@mit.edu}
\author[inst1]{L.~Capozza}
\author[inst3]{J.~Diefenbach\fnref{fn1}}
\fntext[fn1]{Currently with Johannes Gutenberg-Universit\"at, Mainz, Germany}
\author[inst1]{B.Gl\"aser}
\author[inst1]{Y.~Ma\fnref{fn2}}
\fntext[fn2]{Currently with Advanced Meson Laboratory, Nishina Centre, RIKEN, Japan}
\author[inst1]{F.E.~Maas}
\author[inst1]{D. Rodr\'{i}guez Pi\~{n}eiro}

\address[inst1]{Johannes Gutenberg-Universit{\"a}t, Mainz, Germany}
\address[inst2]{Massachusetts Institute of Technology, Cambridge, MA, USA}
\address[inst3]{Hampton University, Hampton, VA, USA}

\begin{abstract}
 
Precise luminosity measurements for the OLYMPUS two-photon exchange experiment at DESY were performed by counting scattering events with alternating beams of electrons and positrons incident on atomic electrons in a gaseous hydrogen target.  Final products of M{\o}ller, Bhabha, and pair annihilation interactions were observed using a pair of lead fluoride Cherenkov calorimeters with custom housings and electronics, adapted from a system used by the A4 parity violation experiment at MAMI.  This paper describes the design, calibration, and operation of these detectors. An explanation of the Monte Carlo methods used to simulate the physical processes involved both at the scattering vertices and in the detector apparatus is also included.

\end{abstract}


\end{frontmatter}


\section{Introduction}
\label{sec:introduction}

\subsection{Purpose}
\label{sec:purpose}

A significant discrepancy persists between the results of two classes of experiments that have measured the ratio of proton form factors, $G_E/G_M$~\citep{Guichon:2003}.  The gap between data from polarization-transfer experiments and those from analyses using Rosenbluth separation could be due in part to contributions from two-photon exchange. Observable effects of this proposed explanation have been modeled in various ways~\citep{Arrington:2011} and a precise measurement is needed to test the validity of published theories.  The OLYMPUS experiment~\citep{Milner:2014} aims to quantify two-photon exchange over a range of four-momentum transfer of 0.4~GeV$^2$/$c^2<Q^2<2.2$~GeV$^2$/$c^2$ through a measurement of the ratio of electron--proton to positron--proton elastic scattering cross sections with a total uncertainty of less than $1\%$.

Installed at the DORIS storage ring at DESY, OLYMPUS collected data with a circulated 2.01~GeV lepton beam, alternating about daily between electrons and positrons, incident on a gaseous hydrogen target.  Kinematics were overconstrained by coincident detection of the scattered lepton and the recoiling proton in a large-acceptance spectrometer inheriting from that used in the BLAST experiment at MIT-Bates~\citep{Hasell:2009}.  Crucial to achieving the desired accuracy of the results is a precise measurement of the luminosity.  While precision on an absolute scale is always desirable, for a result that relies on a ratio it is the precision of the relative luminosity measurement between beam species that is of prime concern.

To this end, multiple subsystems were used to make complementary luminosity determinations based on distinct physical signals.  First, slow control software provided a quick estimate based on the temperature of the target cell, the flow rate of hydrogen into the cell, and the beam current.  Second, small-acceptance tracking detectors at polar scattering angles of about 12$^{\circ}$ were used to count events in that region, where the lepton--proton elastic scattering cross-section is higher than in the spectrometer and where two-photon exchange effects are expected to be negligible~\citep{Arrington:2011} .  Finally, interactions between the beam and atomic electrons in the target were monitored with a pair of Cherenkov calorimeters placed about 3~m downstream of the target center at the symmetric M{\o}ller scattering angle, that is, the polar scattering angle in the lab frame common to both outgoing electrons when they have the same energy.  For a beam energy of 2.01~GeV incident on a stationary target, this is 1.29$^{\circ}$.

This paper describes the design and operation of these ``Symmetric M{\o}ller/Bhabha'' (SYMB) calorimeters.  They were built to handle an accepted rate of $e^{\pm}$--$e^-$ interactions (with products detected in coincidence) of typically 5~kHz.  The choice of detector material needed to be sufficiently radiation hard to maintain consistent performance levels while absorbing a significant dose of high-energy particles and able to provide a fast physical response allowing for signal collection at a kilohertz rate. These considerations led to the use of PbF$_2$ crystals, in which energetic charged particles produce Cherenkov light with no scintillation component.  Photomultiplier tubes (PMTs) gathered the light from each detector.  Their output was passed through analog-to-digital converters (ADCs) and recorded on 8-bit by 8-bit two-dimensional histogramming cards.


\subsection{Theoretical considerations}
\label{sec:theory}

\begin{figure}[t]
	\centering
	\includegraphics[width=0.45\textwidth]{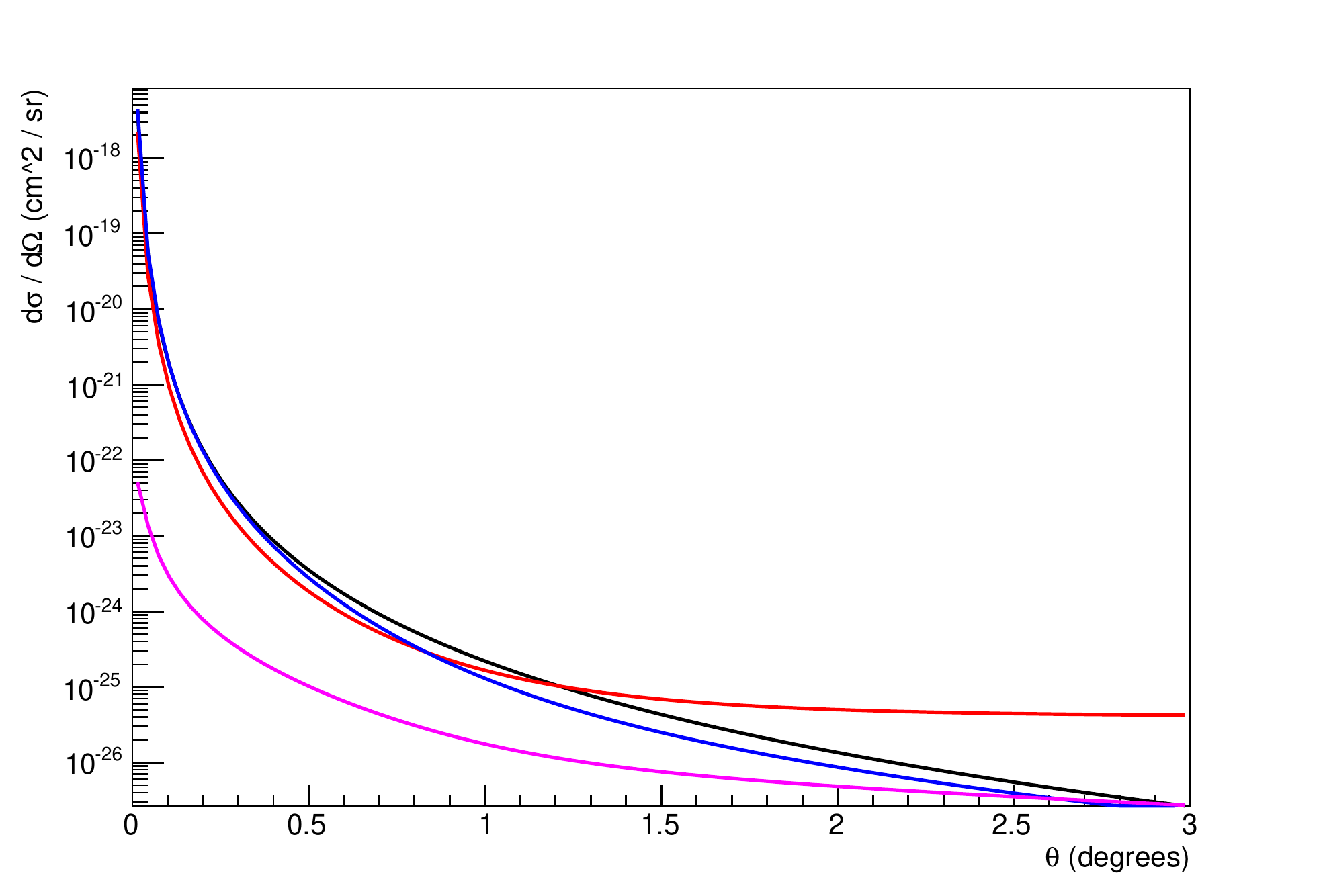}
	\caption{Elastic cross sections as a function of lab-frame scattering angle: M{\o}ller scattering (red), Bhabha scattering (blue), pair annihilation (violet), and electron-proton scattering (black).}
	\label{fig:crosssections}
\end{figure}

Three elastic processes contributed to the observed physical signal: M{\o}ller scattering ($e^- e^-\rightarrow e^- e^-$), Bhabha scattering ($e^+ e^-\rightarrow e^+ e^-$), and pair annihilation ($e^+ e^-\rightarrow \gamma \gamma$).  All three are pure quantum electrodynamic processes whose cross sections are depicted around the region of interest in \cref{fig:crosssections} along with the electron--proton scattering cross section in the Born approximation (where it is equivalent to the positron--proton scattering cross section) for reference.  Next-to-leading order corrections~\citep{Tsai:1960}, including corrections due to radiative final states, were accounted for in the analysis by means of Monte Carlo simulation. The simple nature of these quantum interactions, along with the high acceptance-integrated cross-sections relative to those of elastic $e^{\pm}$--$p$ scattering, provided an opportunity for very precise luminosity measurements at OLYMPUS.

\section{Design}
\label{sec:design}

\begin{figure*}[t]
	\centering
	\includegraphics[width=\textwidth]{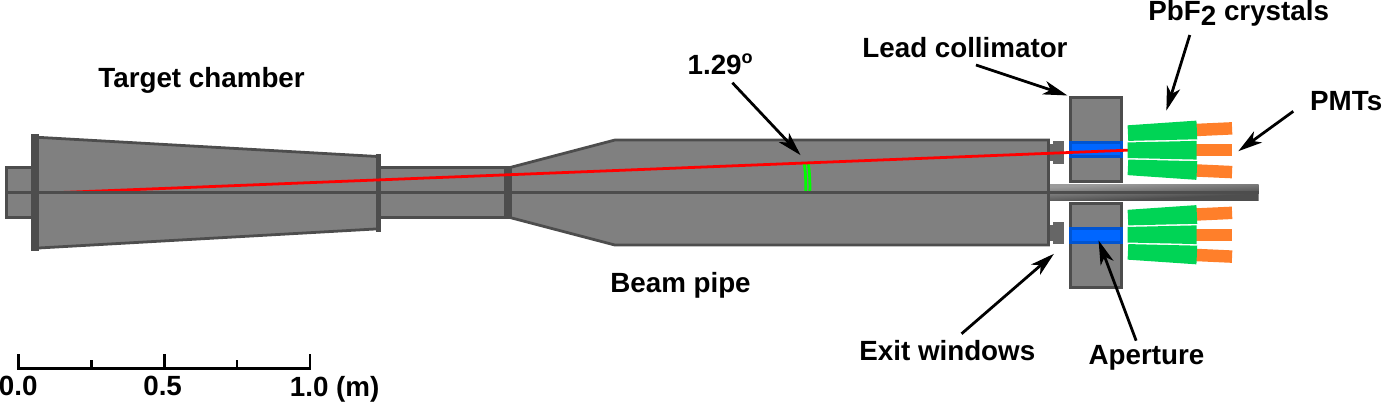}
	\caption{A scale representation of the critical elements.  The support table and other electronics are omitted for clarity and other OLYMPUS detector systems are not drawn.}
	\label{fig:schematic}
\end{figure*}

Two identical SYMB detectors were built in Mainz, one for each ``sector'' of OLYMPUS (the left and right sides, from the beam's perspective).  Each detector consisted of a 3$\times$3 array of lead fluoride (PbF$_2$) crystals placed inside a mu-metal box along with a PMT for each crystal and voltage dividers.  Just outside the box, on the side facing the target, a lead (Pb) collimator was installed.  All components were fastened to a support table so that they could be moved away from the beam line in a controlled way in order to avoid radiation damage during DORIS injections.  They could then be precisely returned back to their original positions.  In practice, injections were well controlled and this functionality was not needed during normal operation.  A schematic of the SYMB detector, target, and beam pipe is presented in \cref{fig:schematic} to provide a sense of scale. 


\subsection{PbF$_2$ crystals}
\label{sec:PbF2}

All crystals used in the detectors were provided by the A4 collaboration at MAMI in Mainz \citep{Baunack:2009}.  Note that these were not regenerated crystals previously used by A4, but ``brand new'' unused crystals.  The lengths of the crystals varied from 150.0~mm to 185.4~mm, each tapered slightly from its front (upstream) face to its back (downstream) face.  These faces were trapezoids, with an area of $\sim670$ mm$^2$ for the front and $\sim900$ mm$^2$ for the back.  Given PbF$_2$'s radiation length of 9.34~mm and Moli\`{e}re radius of 21.24~mm \citep{Beringer:2012}, such dimensions are sufficient for a 3$\times$3 array to contain more than 95\% of the energy of an electromagnetic cascade \citep{Achenbach:2001}.


OLYMPUS was designed with an integrated luminosity goal of 4~fb$^{-1}$.  Integrating over the region allowed by the collimator aperture, the accepted cross section was expected to be less than 50~nb for each SYMB process.  With less than 1~GeV deposited in each central crystal per event, and given the size and density of the crystals, the total absorbed dose due to signal events over the course of the experiment was estimated to be no greater than 25~Gy.  Allowing for some additional ionizing radiation from other sources, this was still considered safe, as even 100~Gy would be likely to cause only minor damage to the transmittance of the crystals \citep{Achenbach:1998}.

Since PbF$_2$ is a pure Cherenkov material with no slow component in its light output, it has a fast rise time of $\sim$5~ns and the full pulse width is well contained within a 20~ns window.  (This was a useful feature for gating signals, since the DORIS beam structure consisted of lepton bunches 24~ps wide and about 100~ns apart.)  Each crystal was wrapped with Millipore paper (Immobilon-P) to improve internal reflection at the faces, then glued to a PMT (Philips XP2900/01).  The custom-made PMT bases were actively stabilized to handle particle rates up to several MHz without any change in gain \citep{Kobis:1998di}.  The completed arrays were tightly bound together by foil and tape.  


\subsection{Collimator}
\label{sec:collimator}

\begin{figure}[t!]
	\centering
	\includegraphics[width=0.45\textwidth]{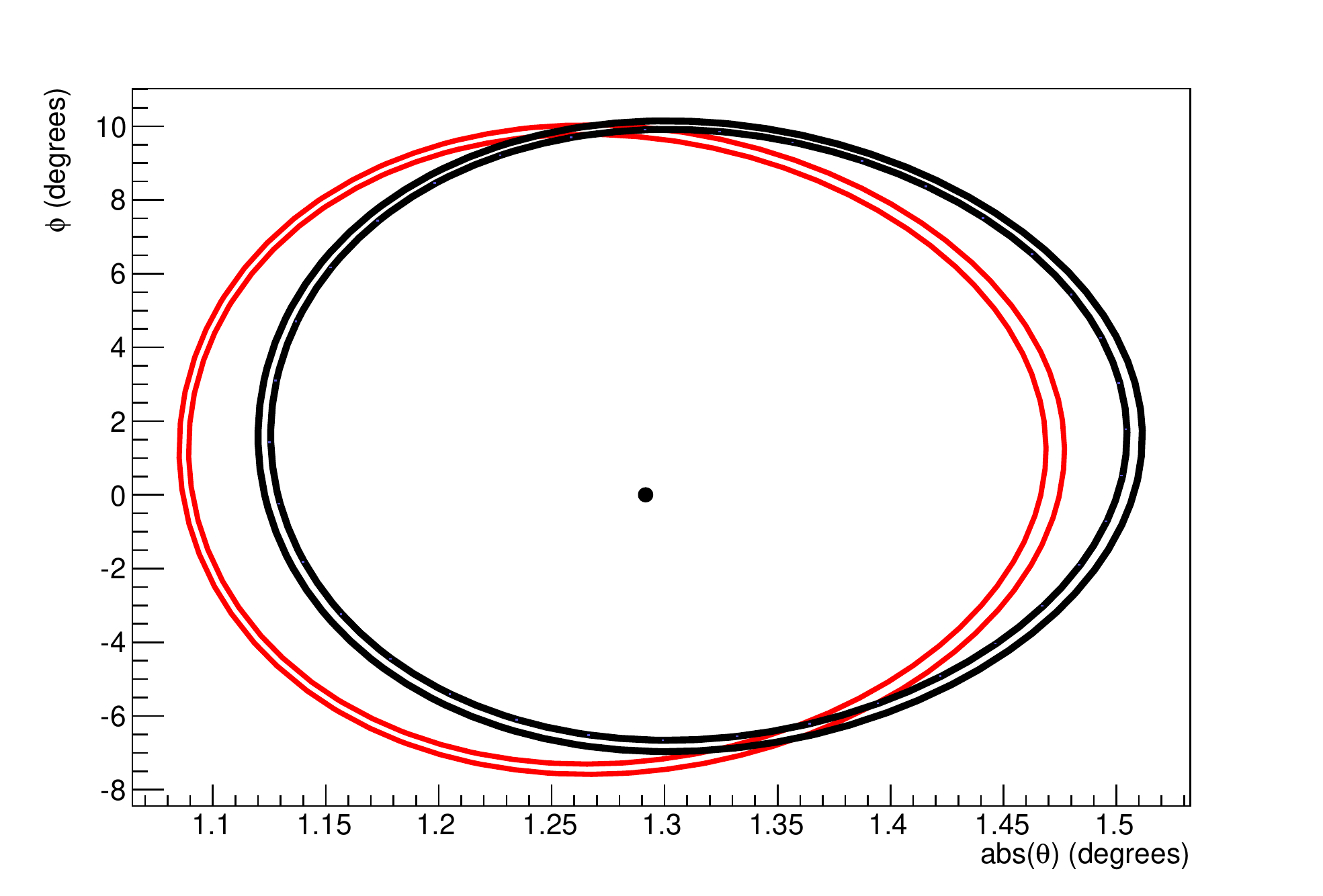}
	\caption{Geometric acceptance of the collimator aperture in the left (black) and right (red) sector.  The outer and inner contour in each pair correspond to the entrance and exit of the collimator, respectively.  A black dot indicates the symmetric scattering angle for M{\o}ller and Bhabha events given a 2.01~GeV beam.}
	\label{fig:acceptance}
\end{figure}

Beam halo and bremsstrahlung prompted the use of a Pb block collimator shielding the front of each detector.  The collimator's dimensions, 200~mm~$\times$~100~mm~$\times$~120~mm, were optimized using Monte Carlo simulation studies of the bremsstrahlung background from the beam pipe.  A cylindrical aperture through the Pb block, 20.5~mm in diameter, was aligned to be coaxial with the long axis of the central crystal in the array.  The strictly geometric acceptances of these apertures relative to the center of the target cell are portrayed in \cref{fig:acceptance}. The left and right sectors have different acceptance due to imperfect alignment of the collimators.

Collimators were constructed out of two pieces: the main bulk of the brick, and a small Pb insert piece that served as the interior (cylindrical) face of the aperture.  This approach was chosen so that the dimensions of the aperture could be decreased by using a different insert if the background rate was found to be too high.

The thickness of each collimator block, 100~mm, would normally be sufficient to contain the energy of a 1~GeV lepton or photon incident on its face.  However, a primary particle that passed partially through the aperture before impinging upon its interior face could shower through a reduced thickness of Pb, and a primary particle that struck the front of the block near the aperture could produce a shower with secondaries reaching the aperture and escaping through it like a tunnel.  These types of events were recorded as data if the resulting energy deposits in the crystals passed the cuts in place; see \cref{sec:operation}.

\subsection{Mu-metal box and driving unit}

OLYMPUS used eight copper coils to generate a toroidal field of about 0.28~T in the region of the tracking detectors.  As a result of imperfect coil alignment, a non-uniform residual field on the order of 10$^{-3}$~T existed in the region SYMB products passed through to reach the calorimeters, as well as at the detectors themselves. \textit{In situ} measurements of the magnetic field allowed for the production of a grid and interpolation scheme that were incorporated into the Monte Carlo simulation to fully capture the effect on particles in transit from the target cell to the various OLYMPUS detectors.

Due to the limited space available at the desired position relative to the target, shielding the PMTs from the magnetic field was a challenge.  The solution was to house each crystal array, along with associated PMTs and voltage dividers, within a mu-metal box. Simulations demonstrated that a thickness of 3~mm provided sufficient shielding.

The SYMB apparatus tables featured an accurate, remotely driven rail system that allowed the boxes to move between ``measurement position'' at 1.29$^{\circ}$ from the beam axis and ``parking position'' further away from the beam pipe.  Extra Pb shielding was placed around the parking position to prevent radiation from entering the calorimeter through the collimator aperture.  An Arduino Mini board was connected to a nearby computer's USB port to serve as the drive controller \citep{Goebel:2011}.


\subsection{Gain monitor system}
\label{sec:gain}

A light pulser system can be used to calibrate and monitor the gain of a PMT. This approach was found suitable for the SYMB detectors as a means to monitor whether any of the outer crystals' PMTs ever stopped functioning.  For this purpose, a pulser was tested at Mainz University \citep{Rehmann:2011}.  This gain monitoring system was used throughout the OLYMPUS data-taking period.



\subsection{Data acquisition electronics}
\label{sec:electronics}

Electronics from the A4 experiment \citep{Kobis:1998ec} were adapted for the OLYMPUS SYMB readout to allow for fast analog summation of the nine crystals' signals in an array with subsequent digitization and histogramming \citep{Kothe:2008zz}.  The system had an overall dead time of only about 5~ns due to the use of a fast first-in-first-out buffer.  A histogramming rate of 50~MHz was achievable.

\begin{figure*}[t!]
	\centering
	\includegraphics[width=\textwidth]{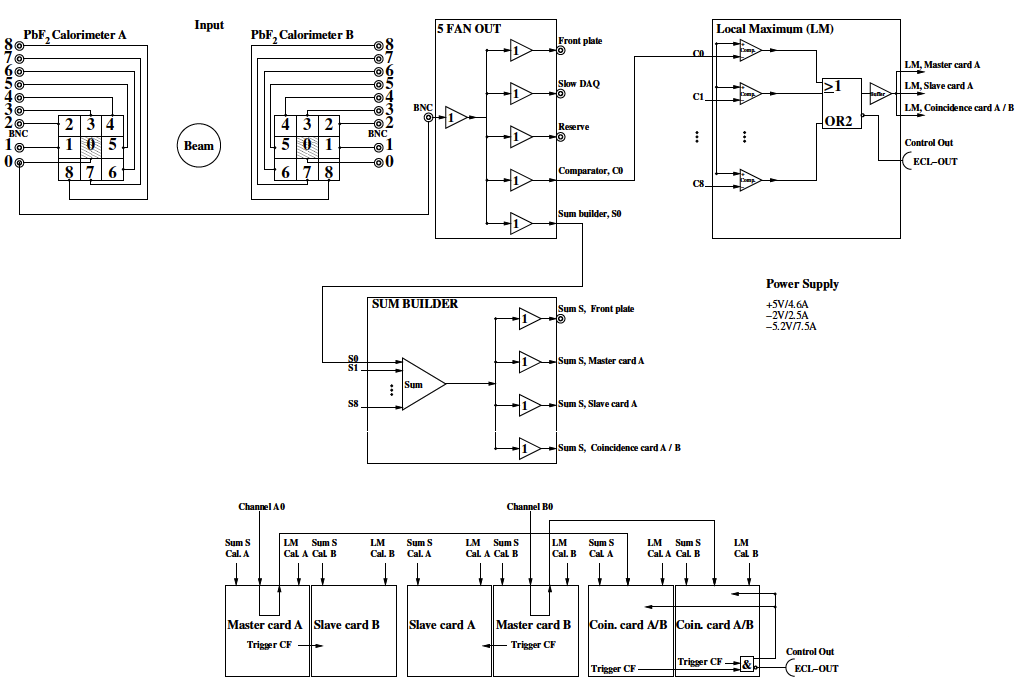}
	\caption{Schematic for the signal flow in the SYMB data acquisition electronics.}
	\label{fig:signalflow}
\end{figure*}

The signal handling is shown in \cref{fig:signalflow}.  First, the nine analog signals (one from each crystal in a sector) are fanned out so that they can be sent to both a sum builder and a ``Local Maximum'' (LM) veto.  The veto accepts an event only if the central crystal in the array had a signal with the highest amplitude, i.e., contained the center of the electromagnetic shower.  The sum of all nine signals is passed through a constant fraction discriminator (CFD) and then sent to three different ADC cards, operated in Left Master, Right Master, and Coincidence modes simultaneously. In the first two modes, the named sector is treated as the master and the opposite sector as the slave: any signal in the master detector that passes the Master LM veto and exceeds the master CFD threshold is recorded, regardless of what is seen in the slave sector.  In coincidence mode, both sectors are required to have synchronous above-threshold signals and to pass their own LM veto.  In this way the LM veto and CFD threshold provide conditions for a trigger, which sends the signals to the histogramming cards for the appropriate mode or modes.  A gate time of 20~ns was chosen to select exactly one bunch in the DORIS beam.


\section{Calibration}
\label{sec:calibration}

\subsection{PMT gains}
\label{sec:pmt}

\begin{figure}[t!]
  \centering
    \includegraphics[width=0.45\textwidth]{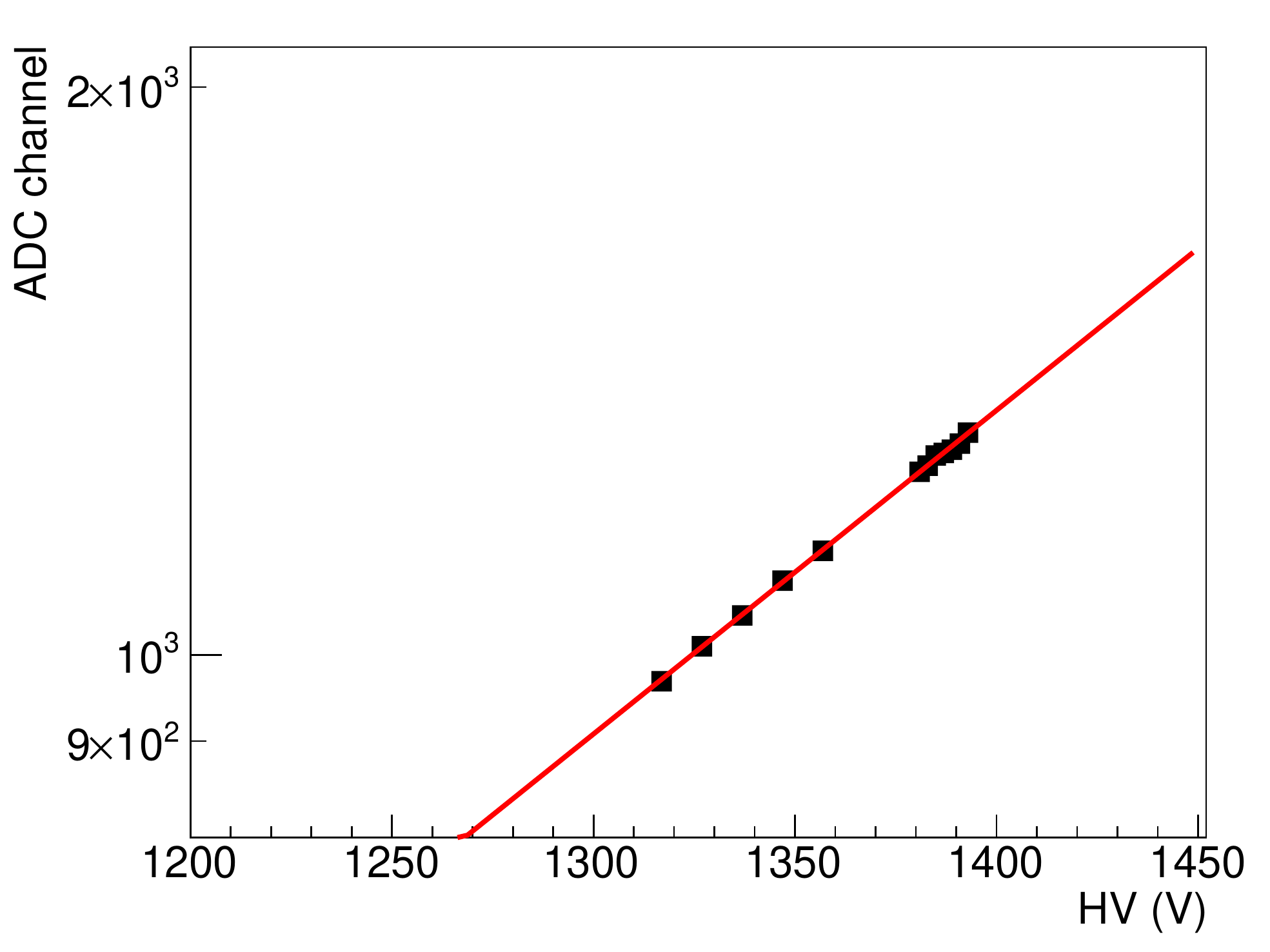}
  \caption{Peak ADC channels versus HV (squares) for crystal \#3 in the left sector and a linear fit (line).}
  \label{fig:hv}
\end{figure}

All PMTs used needed to be calibrated so that a given amount of light would yield the same signal in each of them.  Final gain calibrations were carried out after the PMTs had been glued to the crystals.  The high voltage (HV) of the central crystal in each array was used as a reference, while those of the outer crystals were adjusted to produce the same effective gain as was seen in the central crystal.  Using a 2~GeV positron beam at Test Area 22 at DESY, a series of measurements was taken at different HVs for each of the outer crystals.  A Gaussian filter was applied to the resulting ADC spectra and the output's first derivative was used to find peak positions. For each outer crystal, these peak positions were fitted with a line and the fit results were used to choose an HV setting for the crystal such that the resulting gain matched the gain of the central crystal. \cref{fig:hv} shows a typical example. Since the PMTs were never detached from the crystals, it was assumed that the initial calibration would remain valid over the full OLYMPUS data-taking period.

\subsection{Energy calibration and resolution}
\label{sec:resolution}

Further calibration was required to ensure that the two SYMB calorimeters would provide the same total signal in response to a lepton beam of the same energy.  Using the HV settings from \cref{sec:pmt}, each detector in turn was placed so that the test beam was coaxial with the central crystal while the beam energy was varied in the range of 1--2~GeV in steps of 0.2~GeV. The beam energy spread was constant at 3.2\% during the test beam.

\begin{figure}[t!]
  \centering
    \includegraphics[width=0.45\textwidth]{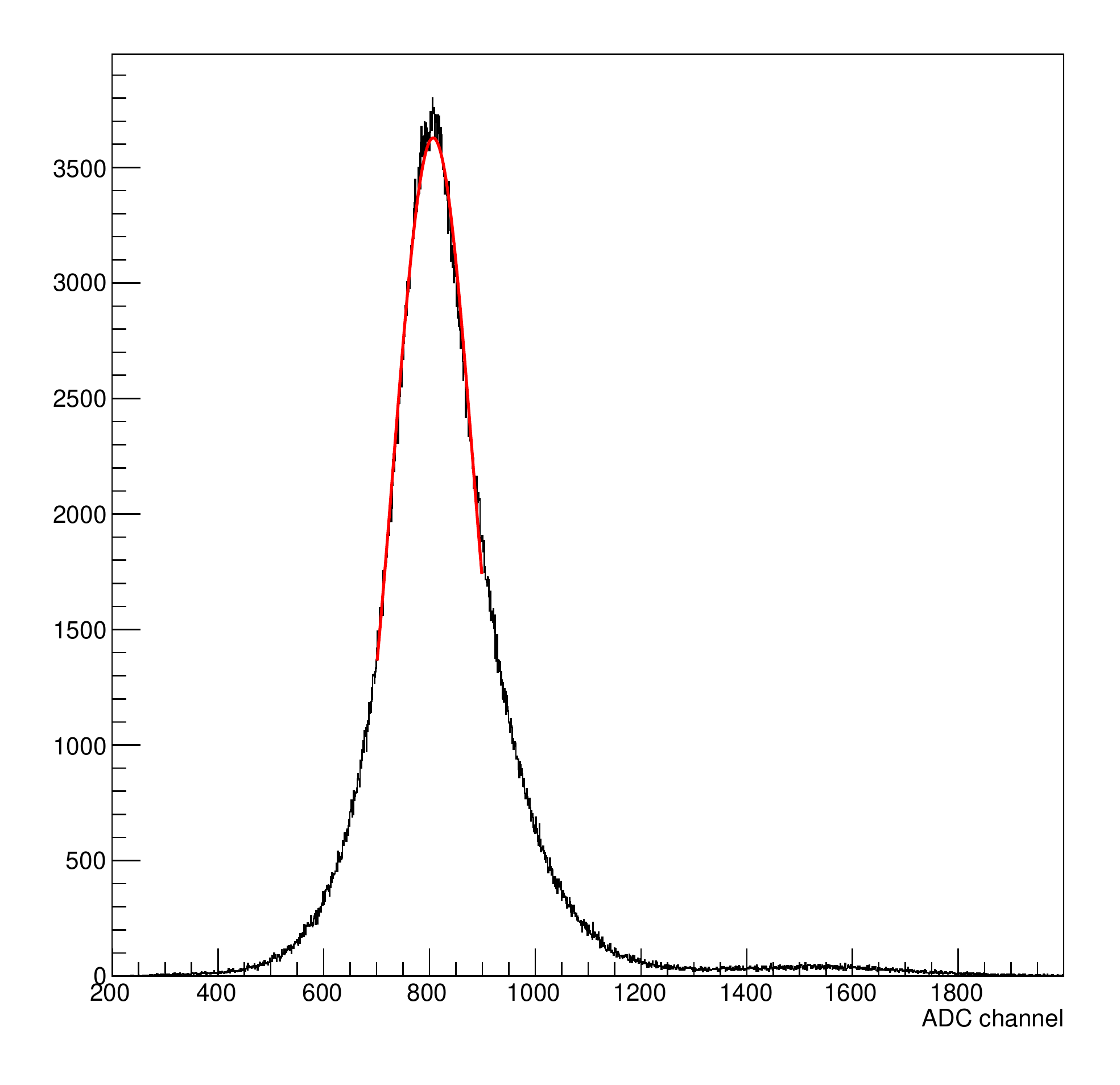}
  \caption{Typical pedestal-corrected ADC spectrum (black) fitted with a Gaussian function (red).}
  \label{fig:box1}
\end{figure}

\begin{figure}[t!]
  \centering
  \includegraphics[width=0.45\textwidth]{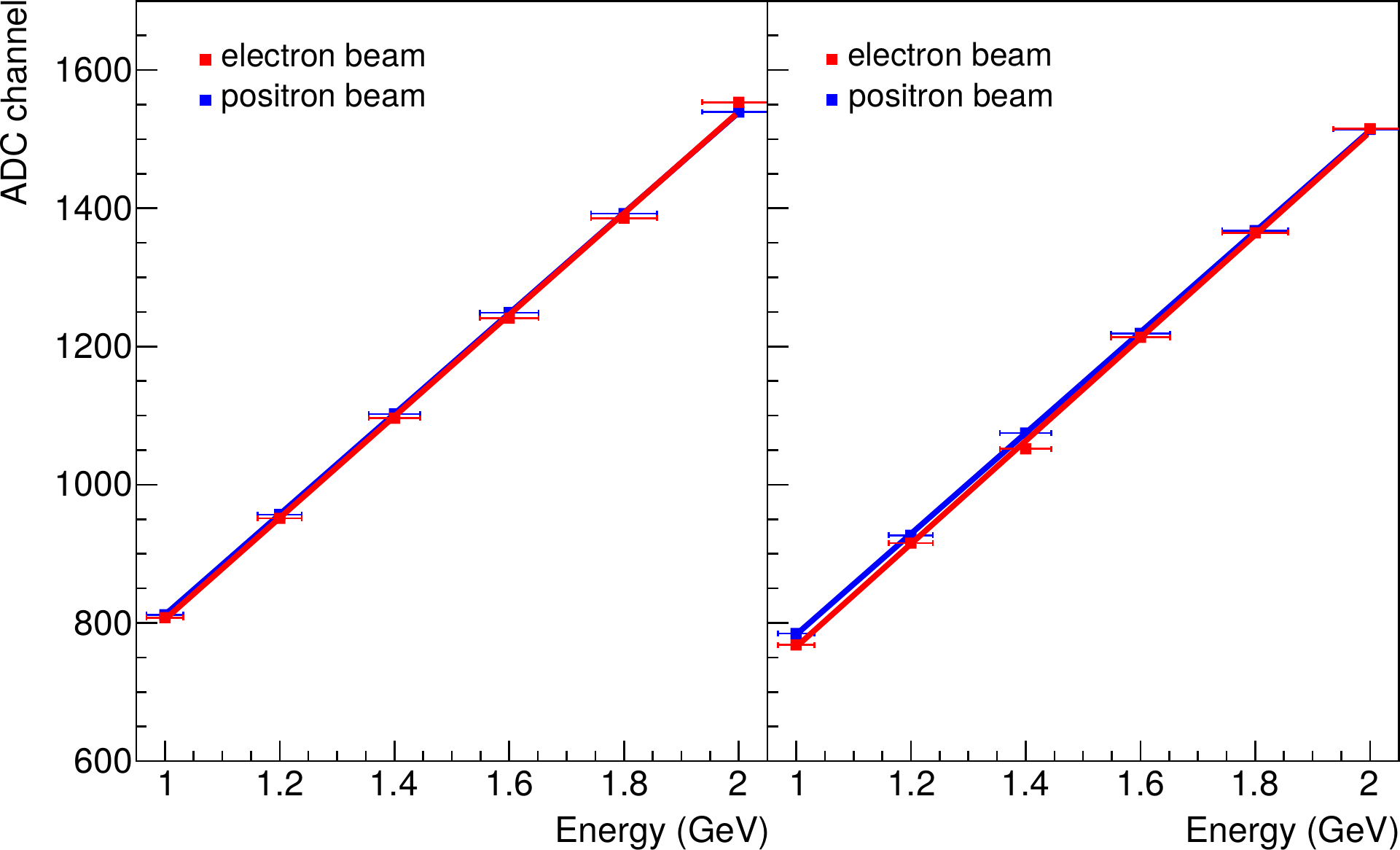}
  \caption{Energy calibration for the left- and right-sector detectors at various beam energies (squares) and fits (lines).}
  \label{fig:box12}
\end{figure}

\cref{fig:box1} shows a typical pedestal-corrected ADC spectrum from the left-sector detector for a 1~GeV electron beam, together with a Gaussian fit.  The second smaller peak around ADC channel 1500 is due to pile-up events.  Using the results of the fit to each spectrum, a relationship between the ADC signal and the beam energy was determined.  \cref{fig:box12} shows the results for both SYMB detectors, together with individual linear fits for electron and positron beams.

\begin{figure}[t!]
  \centering
  \includegraphics[width=0.45\textwidth]{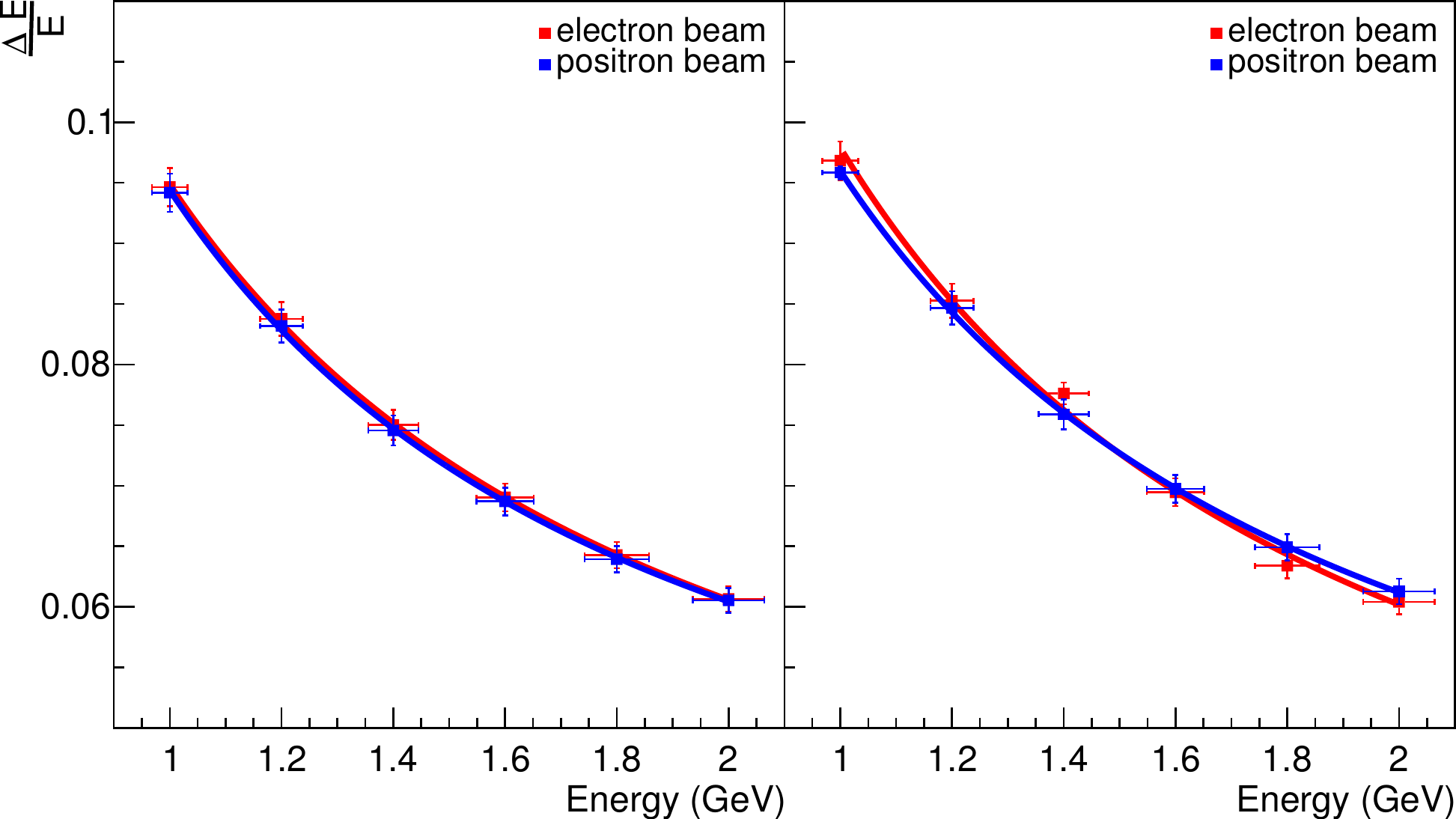}
  \caption{Energy resolution of the left and right detectors at various beam energies (squares) and fits (lines).}
  \label{fig:res12}
\end{figure}

The same data was used for the energy resolution determination, defined as $\Delta E/E=\sigma/\mu$, where $\sigma$ is the standard deviation of the Gaussian fit and $\mu$ is its peak position (see \cref{fig:box1}). The next step was to fit the obtained energy resolution with the following function \citep{Berestetskii:1982, Benisch:2000}:
\begin{equation}
  \label{eq:pmt}
  \frac{\Delta E}{E} = \sqrt{\left(\frac{a}{E}\right)^2 + \left(\frac{b}{\sqrt{E} }\right)^2 + c^2}
\end{equation}
where $a$ represents the electronic noise, $b$ is the statistical fluctuations in the number of detected photons, and $c$ parameterizes the electromagnetic shower fluctuations on the side boundaries of the crystal arrays. \cref{fig:res12} shows the measured energy resolution $\Delta E/E$ of each calorimeter and the fit for both electron and positron beams.  In addition, parameter $a$ was obtained from pedestal measurements.  \cref{tbl:res} summarizes the fit results. The value of $a$ obtained from the fit and from the pedestal measurements ($a$=0.005) are very different, which can be explained by additional contributions from the beam energy assymetry.

\begin{table*}[t!]
  \centering
    \begin{tabular}{ c c c c c } \hline \hline
      Parameter & Left Sector, $e^+$   & Left Sector, $e^-$   & Right Sector, $e^+$  & Right Sector, $e^-$ \\ \hline
      a & $0.070\pm0.037$ & $0.073\pm0.035$ & $0.077\pm0.008$ & $0.071\pm0.035$ \\
      b & $0.057\pm0.021$ & $0.050\pm0.023$ & $0.055\pm0.007$ & $0.057\pm0.020$ \\
      c & $0.029\pm0.006$ & $0.033\pm0.008$ & $0.025\pm0.007$ & $0.029\pm0.005$ \\ \hline \hline
    \end{tabular}
    \caption{Summary of the energy resolution fit results for each SYMB detector and each test beam species.}
    \label{tbl:res}
\end{table*}


\section{Operation, Performance, and Results}
\label{sec:operation}

After calibration, the SYMB detectors were moved into the DORIS beam line together with the whole OLYMPUS apparatus and HV values were set according to the calibration results for optimal detection of SYMB coincidence events.  New readout electronics, utilizing 8-bit ADCs, were used during the data taking.  There were two OLYMPUS run periods, one lasting about a month in early 2012 and another lasting about two months toward the end of that year. In the middle of the first run, HV was decreased to detect elastic lepton--proton scattering products.  Between the first and second runs, HVs were reset to nominal values and four 6~dB attenuators were installed for use in the Master/Slave modes, i.e., each sector's signal was attenuated in each of those two modes.  This effectively doubled the dynamic range of the histogramming cards and successfully allowed for the detection of elastic lepton--proton scattering products throughout the remainder of data taking.

\begin{figure*}[t!]
  \centering
  \subfloat[\label{fig:coincidence}]{\includegraphics[width=0.45\textwidth]{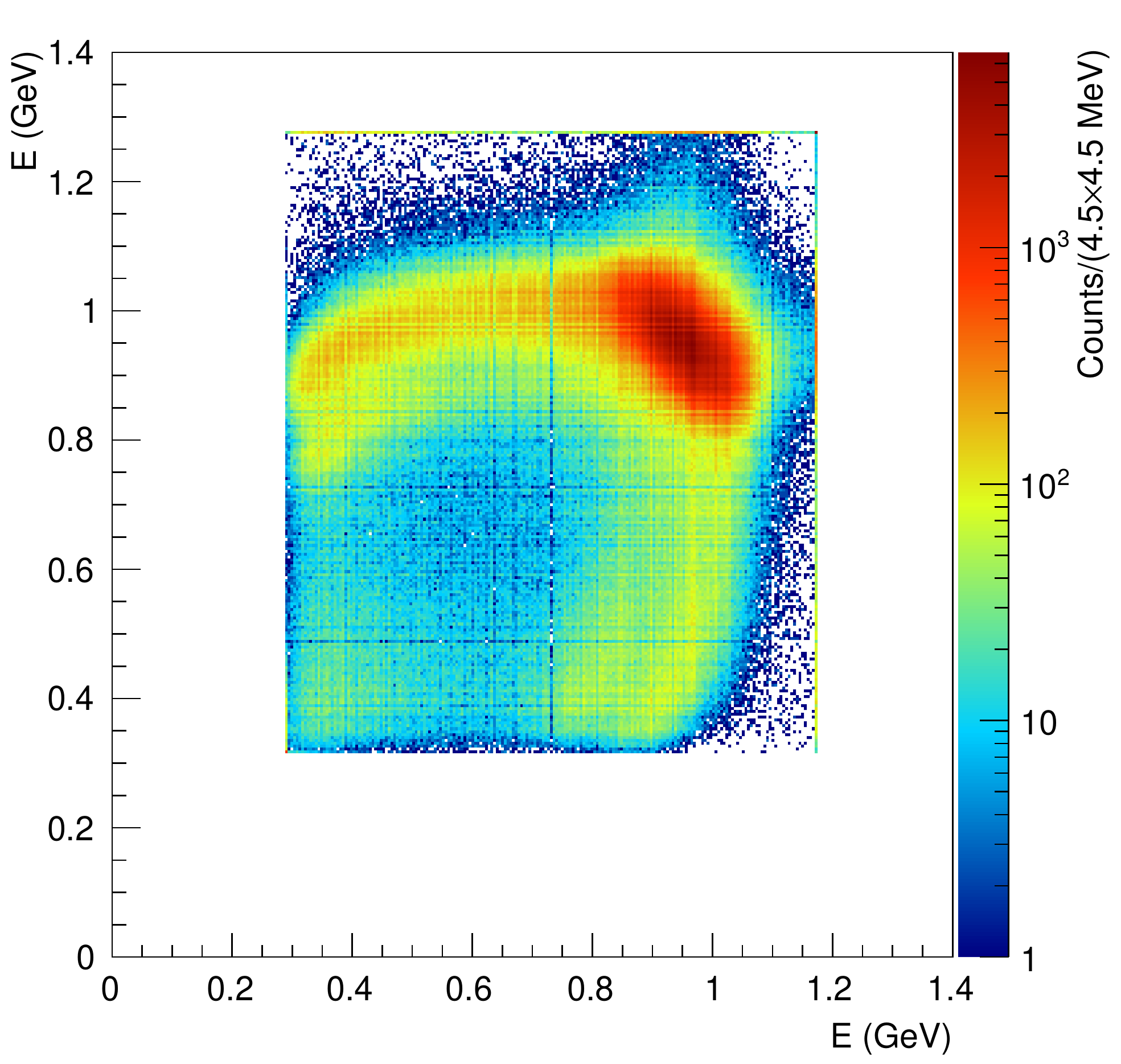} }\hfill
  \subfloat[\label{fig:rightmaster}]{\includegraphics[width=0.45\textwidth]{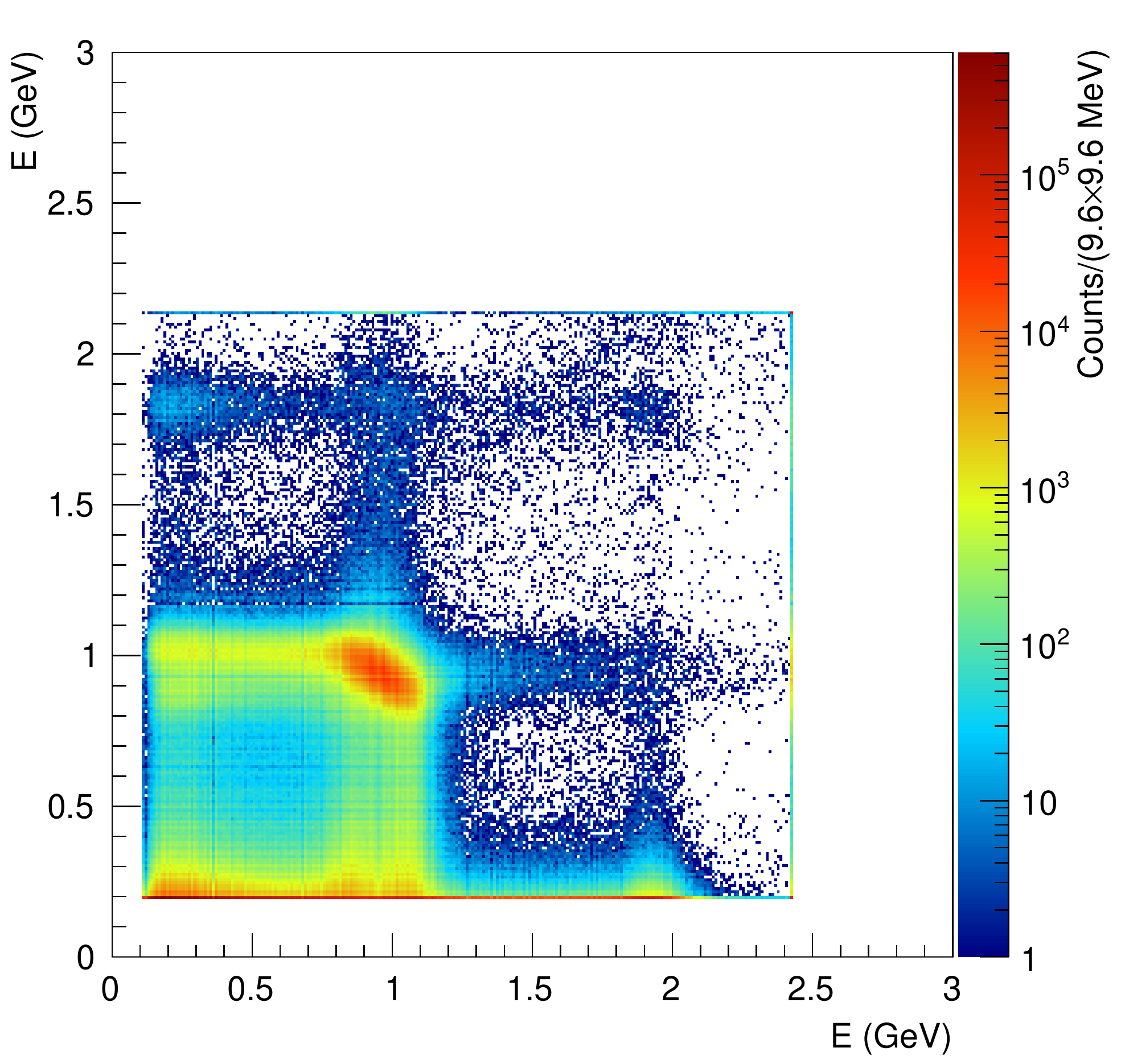} }
  \caption{Typical SYMB data recorded over 18 minutes of beam time by the detector in coincidence mode (a) and right-master left-slave mode (b). The strongest peak in each plot corresponds to coincidence events, while traces of elastic lepton--proton scattering products can be faintly seen as secondary grid lines along the top and right edges of the data in (b).}
  \label{fig:signal}
\end{figure*}

\cref{fig:coincidence} is a typical histogram of the SYMB signal in Coincidence mode. ADC values are plotted for the left sector versus the right sector.  A broad peak which appears as a red oval in the upper right corner represents the coincidence events in which each primary lepton travels directly through a collimator's aperture and deposits most of its energy in the crystals. Horizontal and vertical bands appear due to one particle partially showering through a collimator brick before reaching the crystals while there is a direct hit in the opposite sector associated with the same scattering event. 

The bending of these bands toward the edges of the plot can be explained by the fact that the signal pulse is cut by the discriminator's time gate, consequently decreasing the output signal.  This effect is likely associated with some small time walk and it has no significant impact on the data analysis.  

The entire nonzero region of each plot, with 256 bins along each axis corresponding to the 8-bit ADCs used to collect the data, is offset from the origin due to the negative pedestal values of each ADC, which have been subtracted off.  The colored square region in each plot is therefore indicative of the dynamic range of the ADCs, given their gain settings during the second OLYMPUS run.  The exact positions of the bands and of the oval depend further on physical observables such as the position of the lepton beam.

In order to extract a numerical cross section from one of these two-dimensional histograms, one must choose a region of bins over which to integrate the total counts.  This can be done in several ways, such as fitting an ellipse tightly to the peak region (which would lead to a cross section for the most direct type of coincidences, with no showering in either sector's collimator) or including the entire field other than the underflow and overflow bins (which would lead to a cross section for all coincidences observable in the detectors' dynamic ranges).  In practice, for the sake of OLYMPUS, this choice proves to be free, as the same ratio persists between counts in data and counts in simulation (see \ref{sec:simulation}) as long as the same integration scheme is applied to both.  Yields shown in this report reflect the latter of the options mentioned above.

\cref{fig:rightmaster} shows a typical SYMB signal in Master/Slave mode. In addition to the coincidence events (now in the left bottom quarter of the figure due to a roughly factor-of-two difference in effective gain between Coincidence versus Master/Slave modes), elastic lepton--proton scattering products can be faintly seen as secondary grid lines along the top and right edges of the data.  At their bases, these lines end in more easily visible light-blue peaks.  

Given OLYMPUS's 2.01~GeV beam, each product in SYMB scattering will be roughly 1~GeV.  But for a lepton--proton scatter to produce a particle at 1.29$^{\circ}$, that particle would be close to 2~GeV.  In \cref{fig:rightmaster}, the eye can distinguish two horizontal and two vertical lines, corresponding roughly to these values of energy deposited in either sector.  One can imagine a square formed where these lines intersect: the bottom-left corner of the square is brightly visible as the SYMB coincidence peak, but the other corners are not completely vacant, as these are populated by random coincidences between particles produced at different scattering vertices that arrive at the two calorimeters within the same gate.

There is a visible effect in \cref{fig:coincidence,fig:rightmaster} due to differential nonlinearities in the ADC cards.  The ADC bin widths, in energy, are not all identical, but each bin represents one bit in the digitized output.  Therefore, occasional thin lines (of width one bin) may appear horizontally or vertically that contain slightly more or fewer counts than the surrounding bins.  This is a minor effect, handled in the detailed analysis, but it can give the images the appearance of an aliasing error.

During the first OLYMPUS run in early 2012, measurements from the SYMB calorimeters and other luminosity monitors revealed an observed luminosity that was roughly one-eighth of the expected value.  This led to the discovery of a leak in the gas supply line upstream of the target cell, which was then repaired.  Luminosity measurements subsequently confirmed an increase up to nominal levels.

It is encouraging to demonstrate broad agreement between two independent subsystems designed to determine the luminosity at OLYMPUS.  While the SYMB detectors counted coincidences, the slow control recorded observables such as livetime and estimated the time-dependent luminosity based on DORIS and target parameters.  

\begin{figure}[t!]
  \centering
    \includegraphics[width=0.45\textwidth]{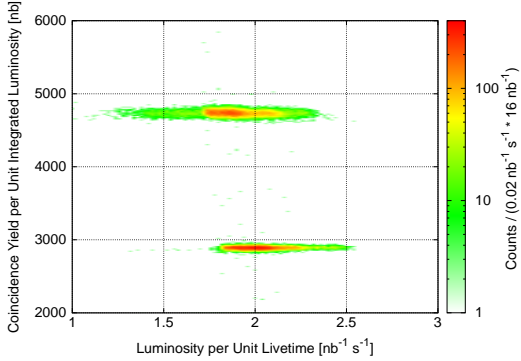}
  \caption{For each beam species, SYMB counts scale linearly with the luminosity estimate from an independent subsystem.}
  \label{fig:current}
\end{figure}

The SYMB detectors integrated coincidence counts over discrete readout periods, whose durations were determined by the rate of lepton--proton scatters counted by the OLYMPUS spectrometer but which typically lasted about half a minute.  Plotting the coincidence yield from the SYMB normalized by the luminosity obtained from the slow control, shown in \cref{fig:current} versus the integrated luminosity in each readout period, reveals two groups of points that each appear as tightly clustered about some constant value (i.e., with no significant dependence on the abscissa).

The upper group corresponds to electron beam running and the lower group to positron beam.  Their best-fit constant values are 4734~nb and 2874~nb respectively, with a ratio of 1.647, agreeing with expectations from estimated accepted cross sections obtained from radiative Monte Carlo simulation to within a few percent.  The lack of any distinct shape or structure diverging from variation around a fixed value suggests that the SYMB detectors were not prone to lose efficiency as counting detectors due to high or low signal rates.

\begin{figure}[t!]
  \centering
    \includegraphics[width=0.45\textwidth]{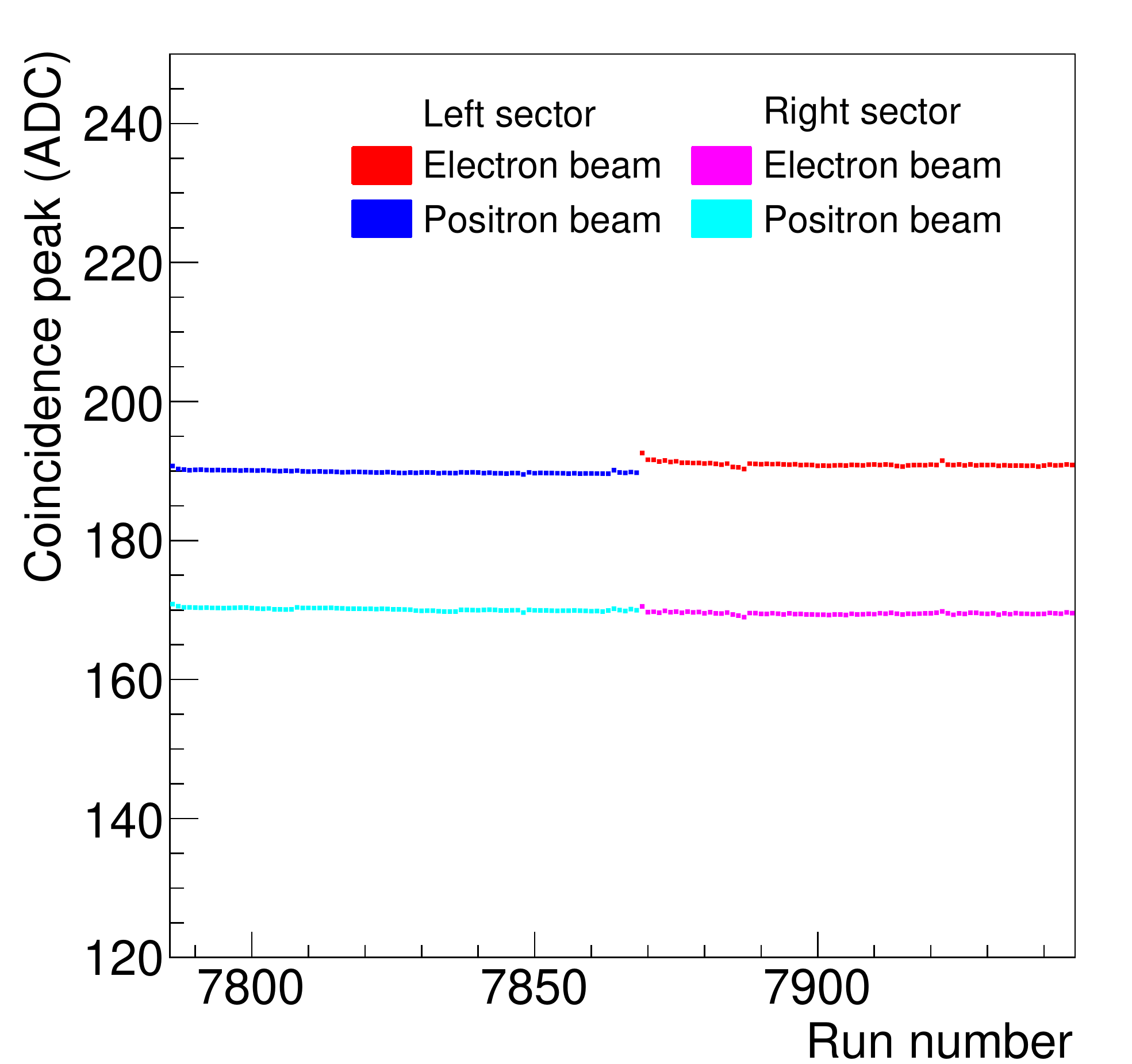}
  \caption{Positions of the coincidence peaks in the left and right sectors for both beam species over about two days of data collection.}
  \label{fig:peak_position}
\end{figure}

\cref{fig:peak_position} shows the stability of the coincidence peaks in the ADC channels over about two days of data collection. The difference between the peak heights in the left and right sectors is due to different gains.  Switching beam species caused peak positions to briefly move up by a few ADC channels, but otherwise they remained very stable over long stretches.

\begin{figure*}[t!]
  \centering
  \subfloat[\label{fig:beamscans_offsetx}]{\includegraphics[width=0.45\textwidth]{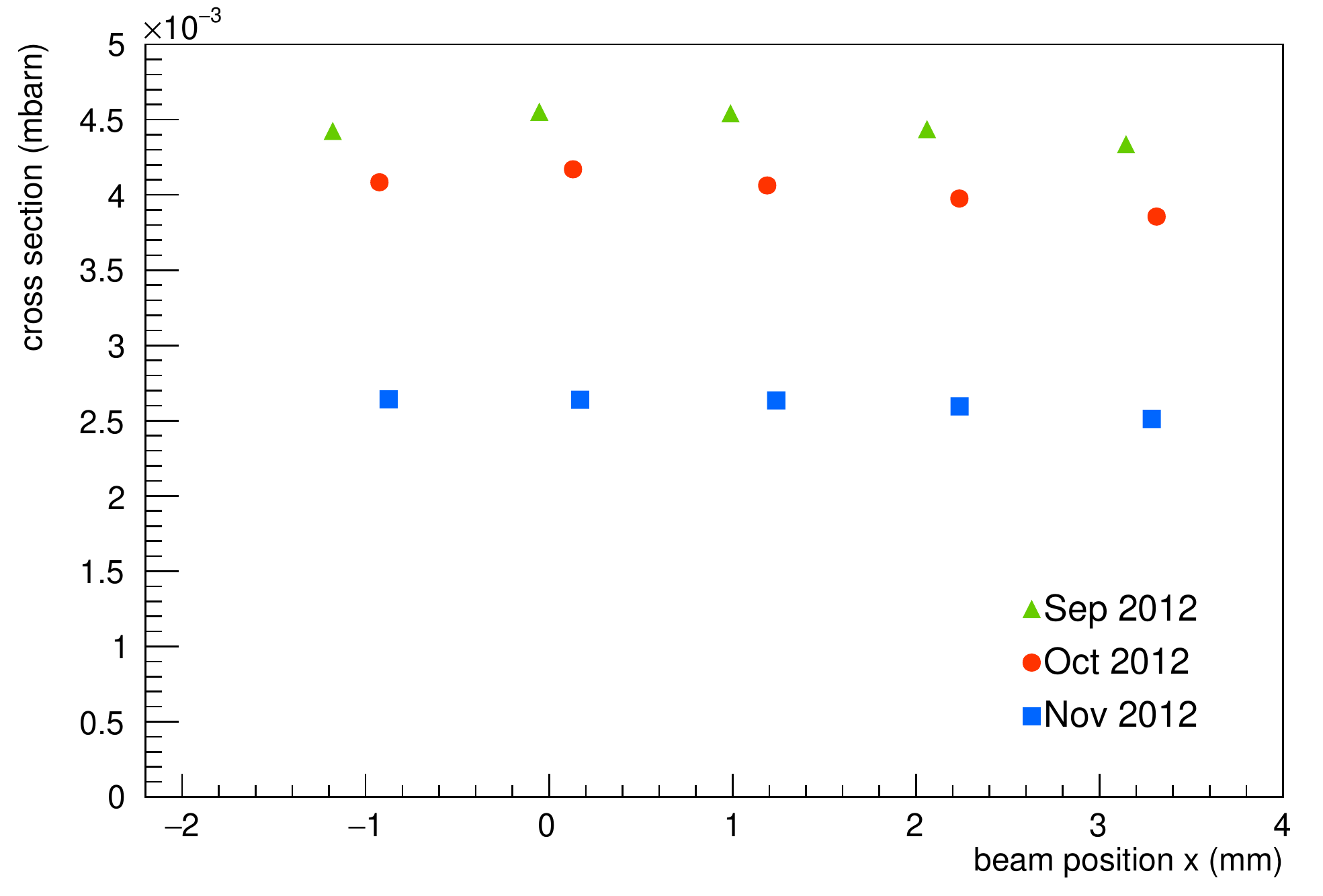} }
  \subfloat[\label{fig:beamscans_offsety}]{\includegraphics[width=0.45\textwidth]{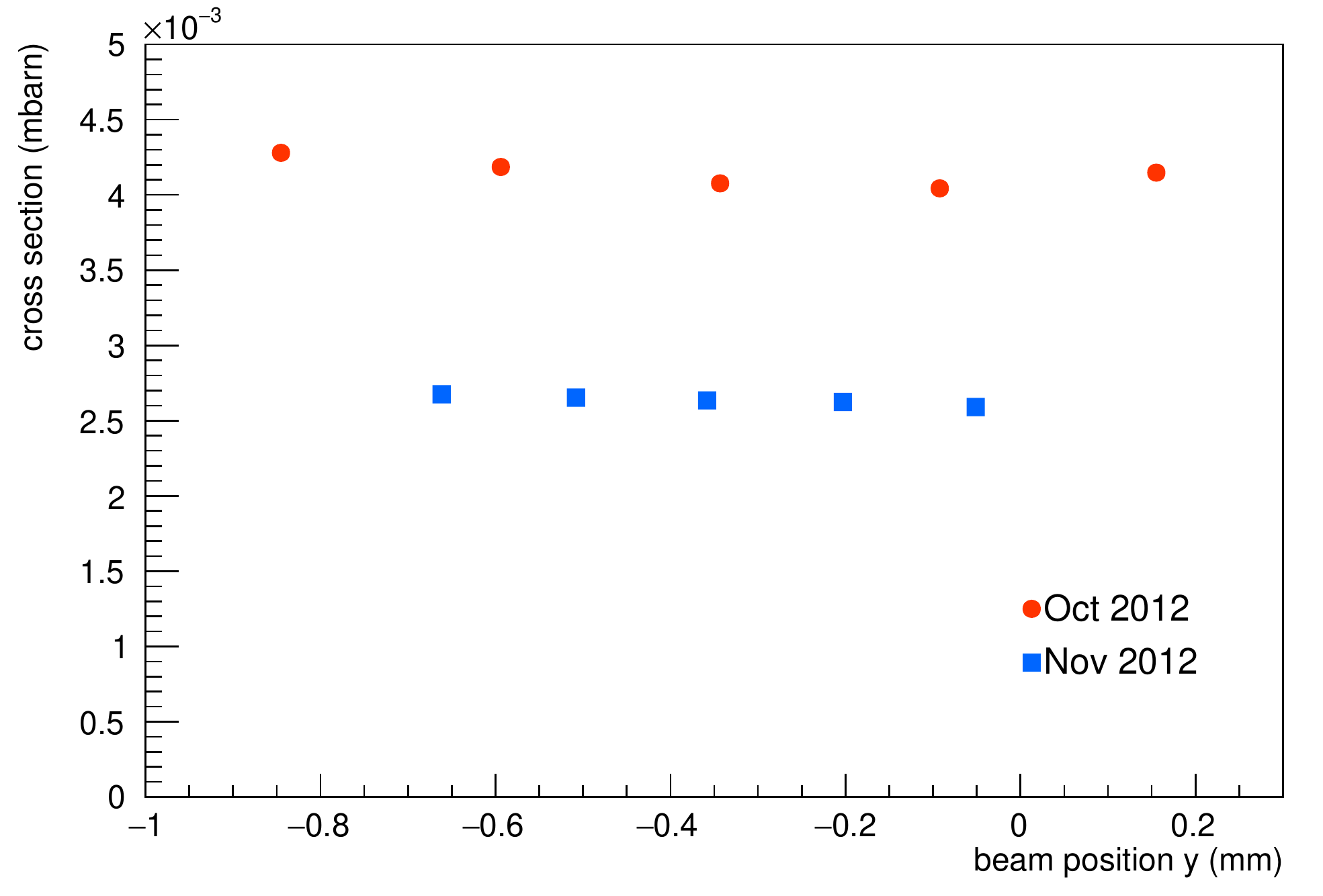} } \\
  \subfloat[\label{fig:beamscans_slopex}]{\includegraphics[width=0.45\textwidth]{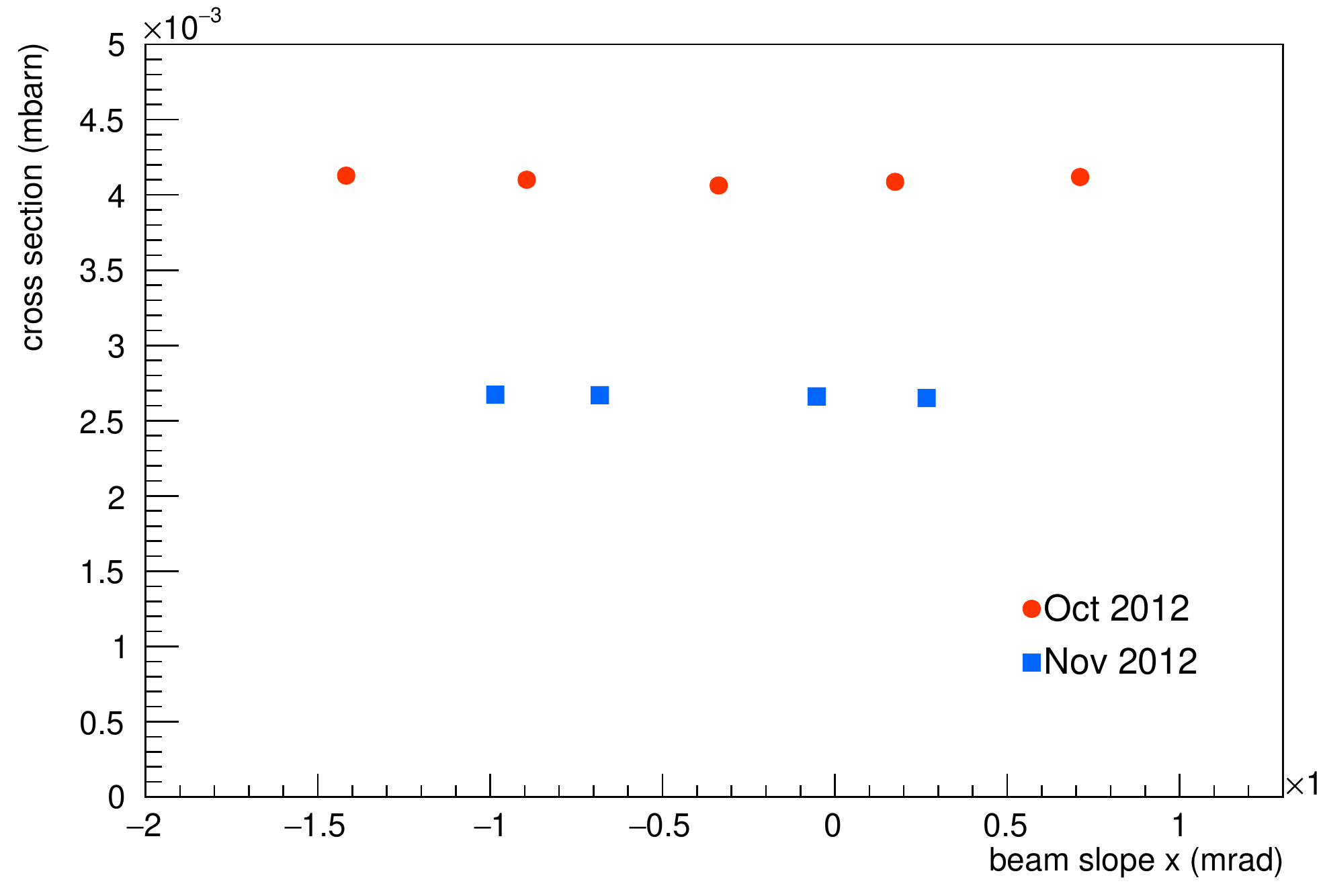} }
  \subfloat[\label{fig:beamscans_slopey}]{\includegraphics[width=0.45\textwidth]{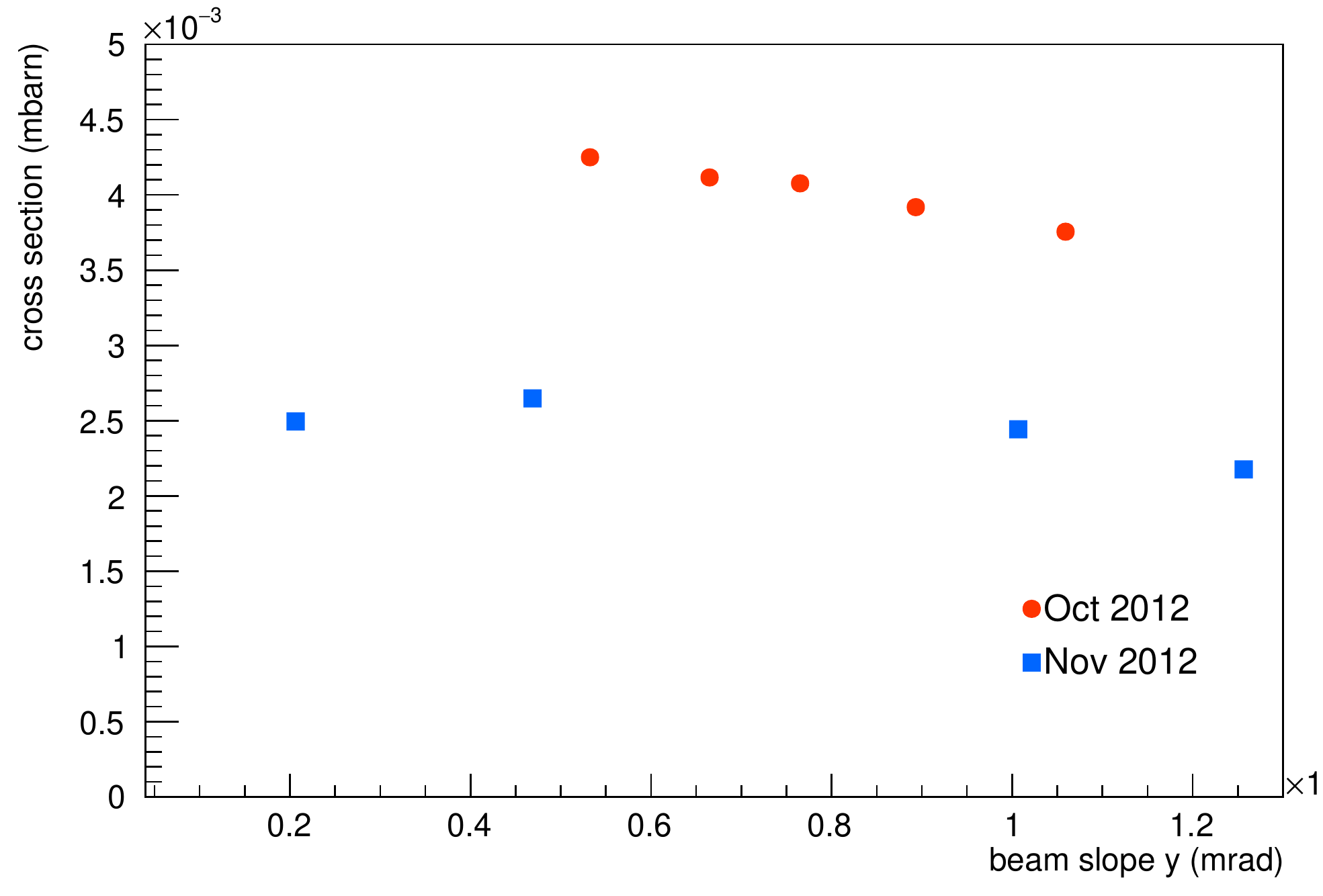} }
  \caption{Count rate in the SYMB during beam position (a, b) and slope (c, d) scans along the horizontal and vertical axes. Most error bars are smaller than markers.}
  \label{fig:beamscans}
\end{figure*}

During the second OLYMPUS run, several beam position and beam slope scans were performed as tests.  In each scan, the DORIS beam's position or slope at the target center was varied along either the vertical or the horizontal axis while remaining fixed along the other axis.  \cref{fig:beamscans} shows cross sections (SYMB coincidence yields normalized by the slow control luminosity) for different beam position and slope scans along the horizontal and vertical axes. These tests demonstrated the sensitivity of the SYMB detectors to beam movements and enabled the optimization of the beam position to maximize SYMB rates.

\section{Simulation}
\label{sec:simulation}

Detailed Monte Carlo studies have been carried out to aid in characterization of the signal, understanding of systematics, and calibration of analysis parameters such as cut placements.  The goal of these studies is to include all relevant physical effects so that the simulated results mimic the data recorded during running.  

The time-dependent luminosity measured by slow control is used as an input to the simulation, although it is assumed to be incorrect by some scale factor on the order of a few percent.  In order for the simulation to match the data exactly, the real correction must be found, whether it is constant or a function of other time-dependent effects.  Comparison of data with simulation will allow for a determination of the proper correction which, in combination with the luminosity measurement from slow control, will provide a definitive measurement of the true luminosity throughout data collection.

These studies utilized the OLYMPUS simulation framework, which is divided into three sequential steps: event generation, particle propagation, and digitization.

\subsection{Event generation and propagation}
\label{sec:generation}

Simple event generators that produce final states (i.e., pairs of particles) for M{\o}ller scattering, Bhabha scattering, and pair annihilation have been written using tree level cross section formulas.  A more advanced set of generators, including next-to-leading order radiative corrections, has also been developed and is intended for use in the full OLYMPUS analysis \citep{Epstein:2016}.  Based on the input beam species and energy, a kinematically allowed result is selected randomly according to a sampling distribution that approximates the cross section as a function of the scattering angle.  The generator assigns the appropriate four-momentum to each outgoing lepton or photon, and their initial position at the event vertex is determined based on the spatial density distribution of the simulated gaseous hydrogen target cell and beam position information that can be artificial or derived from monitor readings in the OLYMPUS data stream.  Each event also receives a numerical weight at generation time, accounting for the sampling distribution as well as the physical cross section of the event's final state, which is used in the final analysis to ensure that counts from simulation can be compared directly with counts in data.

\begin{figure*}[t!]
  \centering
  \subfloat[\label{fig:simulationa}]{\includegraphics[width=0.32\textwidth]{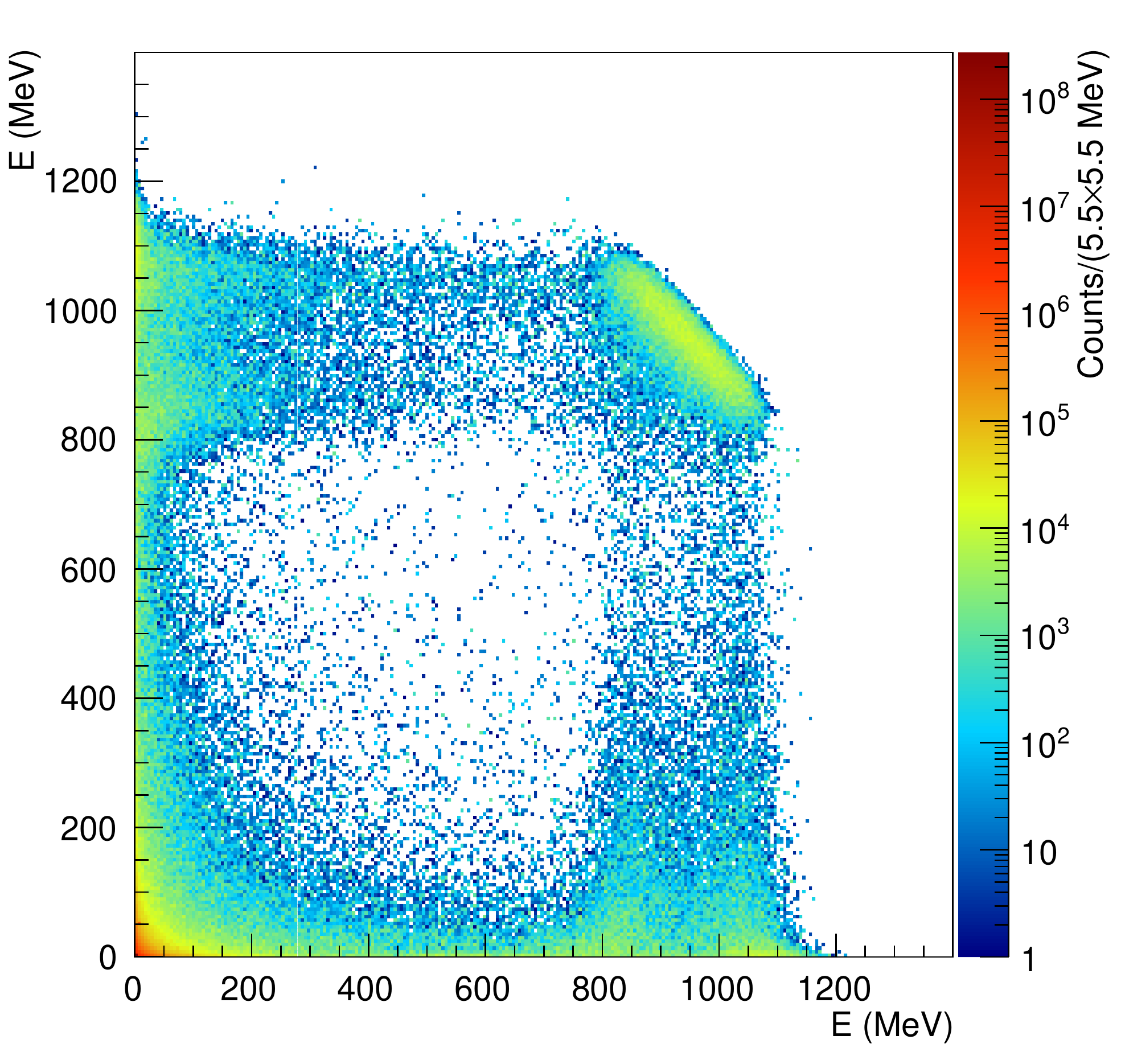} }\hfill
  \subfloat[\label{fig:simulationb}]{\includegraphics[width=0.32\textwidth]{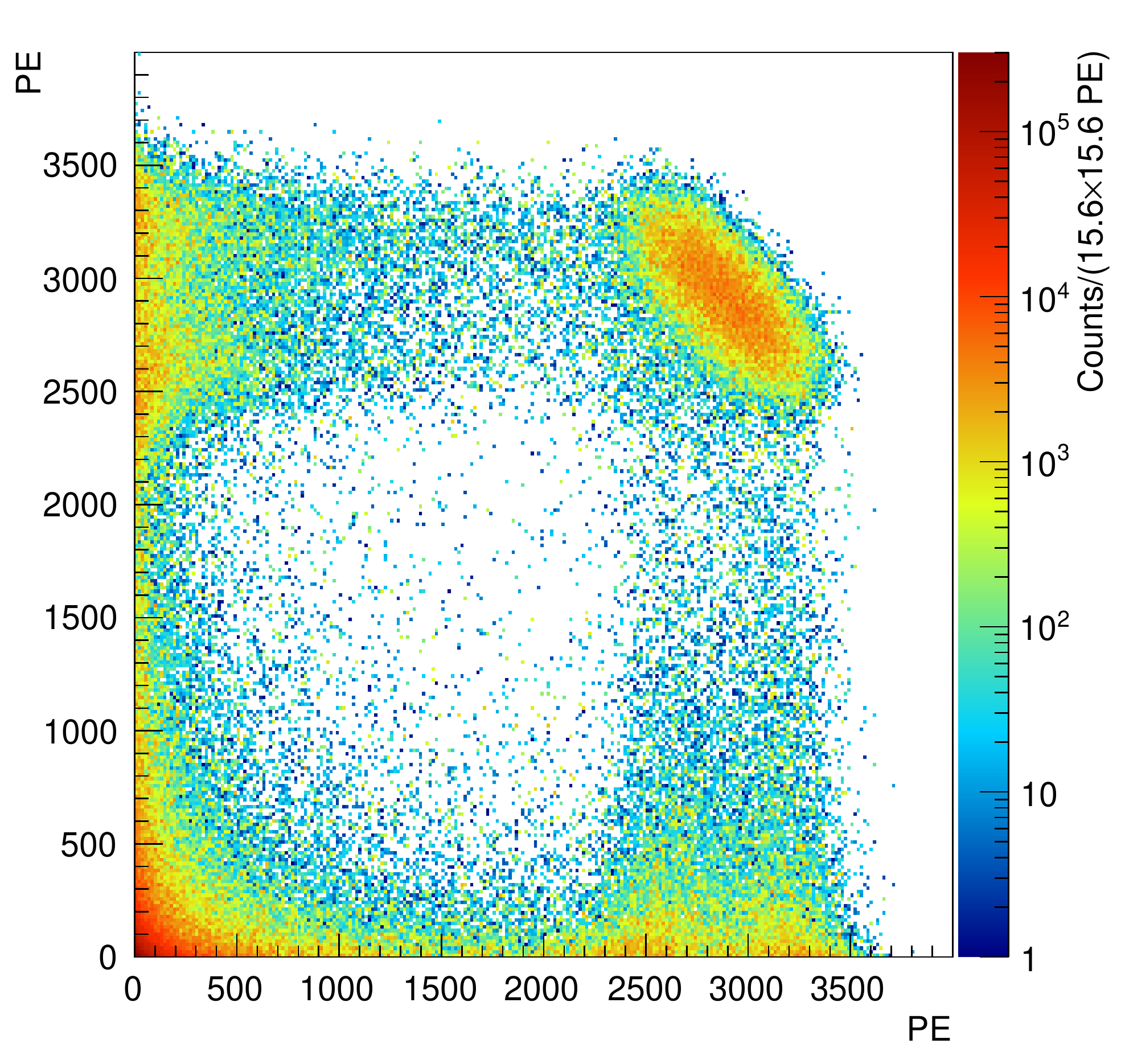} }\hfill
  \subfloat[\label{fig:simulationc}]{\includegraphics[width=0.32\textwidth]{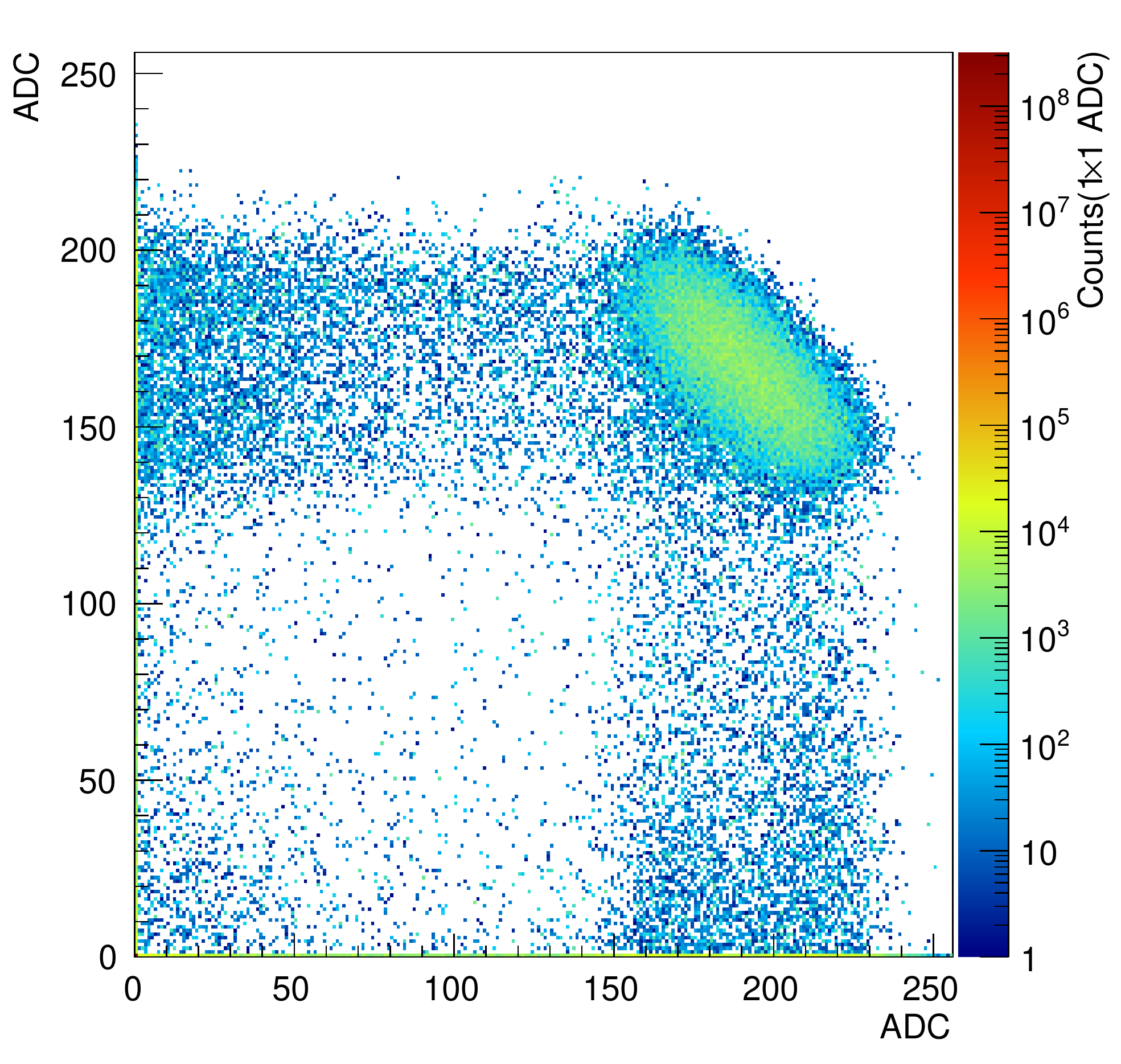} }
  \caption{SYMB digitization process of Monte Carlo data, left sector versus right: (a) energy deposition, (b) number of photoelectrons produced, and (c) ADC signal for M{\o}ller scattering.}
  \label{fig:simulation}
\end{figure*}

Generated particles are propagated through a realistic solid model of the OLYMPUS apparatus. The propagation is carried out using GEANT4 \citep{Agostinelli:2003}.  Detector placements, based on \textit{in situ} surveys, are maintained in the GDML file format \citep{Chytracek:2006}.  Typically, a primary particle's trajectory will proceed down the beam pipe, bending incrementally in the magnetic field, until it passes through the collimator and impinges on the calorimeter, beginning an electromagnetic shower.  Every secondary particle in the cascade is tracked by GEANT4 so that the distribution of energy deposits from ionization are as accurate as possible, with realistic variance.  The energy deposit in each crystal is the output of the propagation step.  \cref{fig:simulationa} plots the sum of deposits in all left-sector crystals versus that in all right-sector crystals at this stage in the simulation process.

\subsection{Digitization}
\label{sec:digitization}

Simulated events are intended to be analyzed in exactly the same way as observed events.  This requires a digitization step, in which data formatted identically to the output of the detector is obtained from the energy deposition produced by GEANT4.

Since PbF$_2$ is a pure Cherenkov calorimeter, one could use the optical methods of GEANT4 to have every charged particle in the showers generate cones of light and to use ray tracing to follow the photons until they either register in a PMT or are lost.  A less computationally expensive approach in two steps has been derived from \citep{Capozza:2010}.  

First, given the total energy loss in a single crystal, an estimate is obtained for the number of photoelectrons produced in the PMT due to Cherenkov light (\cref{fig:simulationb}). This estimate makes use of a parameterization based on data obtained using PbF$_2$ crystals like those in the detector.

Second, a digital ADC signal is determined as a linear function of the number of photoelectrons in the PMT.  The coefficient and scalar offset were obtained from calibration data.  \cref{fig:simulationc} shows ADC signals from Monte Carlo simulation for both the left and right detector modules.

The digitization process accounts for systematic effects and statistical fluctuations due to the associated electronics used in the experiment.  Finally, artificial ADC signals from simulation are recorded in ROOT \citep{Brun:1997} trees with the same structure as the raw OLYMPUS data so that both can be analyzed with the same software.

Because of effects from a small, uncalibrated time walk in the data acquisition electronics that manifest in the raw SYMB data (see \cref{sec:operation}), it is not possible for the simulated results to perfectly match the empirical results.  However, an accurate simulation allows for signal and noise to be distinguished in the data by comparison to a noise-free result from Monte Carlo, and using the same approach to simulate both the electron beam and the positron beam aids in identifying any charge-asymmetric effects in the analysis.  In order to produce a precise relative luminosity measurement, it is crucial to prevent significant systematic effects from data selection techniques that treat M{\o}ller events and Bhabha events on unequal footing.

Also worth noting is that the ratio between counts in data and counts in simulation is consistent regardless of the cuts used, as long as the same scheme is applied in both cases.  This essentially rules out any bin-to-bin variation in systematic effects or inefficiencies in the data.


\section{Conclusion}
\label{sec:conclusion}

The design of a luminosity monitoring system, consisting of a pair of Cherenkov electromagnetic calorimeters, has been presented along with a detailed explanation of their use as luminosity monitors in the OLYMPUS experiment.  From the data obtained using these detectors, integrated rates of lepton--lepton scattering events are being determined now as a function of time over both full OLYMPUS running periods.  Such empirical results can be compared to theoretical expectations by means of a full simulation that accounts for beam dynamics, first-order and radiative scattering processes, and the geometry of the apparatus.  Relative luminosity values can thus be obtained with tightly constrained systematic variance and high statistics, providing an adequate normalization for the OLYMPUS experiment's precision measurement of the electron--proton to positron--proton elastic scattering cross section ratio.


\section{Acknowledgments}
\label{sec:acknowledgments}

This work is indebted to the A4 collaboration for their thorough studies of PbF$_2$ crystals as Cherenkov calorimeters. We are thankful to the electronics workshop from Mainz University's Institut f{\"u}r Kernphysik (IKPH), particularly Mr.~Klein and Dr.~Lauth for discussing the A4 electronics with us at length; as well as the mechanical workshop from IKPH Mainz, especially Mr.~Luzius. We gratefully acknowledge Dr.~Yang for performing simulations of mu-metal boxes. The authors also wish to thank all our collaborators in the OLYMPUS experiment and the staff at DESY, where not only the data collection but most of the calibration of the detectors took place.

This work was supported by the Office of Nuclear Physics of the U.S. Department of Energy, Grant no.~DE-FG02-94ER40818.


\bibliographystyle{elsarticle-num.bst}
\bibliography{SYMB}{}

\end{document}